\documentstyle[12pt,aps]{revtex}
\newcommand{\nn}{\nonumber}
\newcommand{\p}{\partial}
\newcommand{\pr}{\p_r}
\newcommand{\pv}{\p_v} 
\newcommand{\pta}{\p_{\theta}}
\newcommand{\pvi}{\p_{\varphi}}

\newcommand{\spr}{\p_{r_*}}

\newcommand{\spdr}{\p^2_{r_*}}
\newcommand{\spdvr}{\p^2_{r_* v_*}}
\newcommand{\spdra}{\p^2_{r_* \theta_*}}
\newcommand{\spdri}{\p^2_{r_* \varphi}}

\newcommand{\sta}{\sin\theta}
\newcommand{\cta}{\cos\theta}
\newcommand{\sda}{\sin^2\theta}

\newcommand{\coa}{\cot\theta}
\newcommand{\sqd}{\sqrt{2}}
\newcommand{\cD}{\cal D}
\newcommand{\cL}{\cal L}
\newcommand{\cK}{\cal K}
\newcommand{\bL}{\bar{\cal L}}
\newcommand{\drH}{\dot{r}_H}
\newcommand{\ddrH}{\ddot{r}_H}
\newcommand{\prH}{r_H^{\prime}}
\newcommand{\kH}{\kappa_H}
\newcommand{\pprH}{r_H^{\prime\prime}}

\begin{document}
\baselineskip 15pt
\pagestyle{myheadings}
\markboth{}{Wu and Cai}
\draft

\title{\Huge{\bf Hawking Radiation of Dirac Particles \\
in a Variable-mass Kerr Black Hole}}
\author{Wu Shuang-Qing$^*$ and Cai Xu$^{\dagger}$}
\address{Institute of Particle Physics, Hua-Zhong
Normal University, Wuhan 430079, China \\ \rm
$^*$E-mail: sqwu@iopp.ccnu.edu.cn 
~~$^{\dagger}$E-mail: xcai@ccnu.edu.cn}
\date{\today}
\maketitle

\widetext 
\begin{quote}
{\em
Hawking effect of Dirac particles in a variable-mass Kerr space-time is 
investigated by using method of the generalized tortoise coordinate 
transformation. The location and the temperature of event horizon of 
the non-stationary Kerr black hole are derived. It is shown that the 
temperature and the shape of event horizon depend not only on the time 
but also on the polar angle. However, our results demonstrate that the 
Fermi-Dirac spectrum displays a residual term which is absent from that 
of Bose-Einstein distribution.

}
\end{quote} 

\newpage
\narrowtext

An important subject in black hole physics is to reveal thermal properties 
of various black holes. In order to study the Hawking effect\cite{Hawk} of 
a nonstationary black hole, Zhao and Dai\cite{ZD} suggested a novel method 
of the generalized tortoise coordinate transformation (GTCT) which can give 
simultaneously the exact values both of the location and of the temperature 
of the event horizon of a non-stationary black hole. The GTCT method has
been applied to investigate thermal radiation of scalar particles in the 
non-stationary Kerr black hole\cite{ZDH} and in the non-stationary Kerr-Newman 
black hole\cite{LJW}. 

However, it is very difficult to discuss on the evaporation of Dirac 
particles in the non-stationary axisymmetric black hole. The difficulty 
lies in the non-separability of the radial and angular variables for 
Chandrasekhar-Dirac equations\cite{Ch} in the non-stationary axisymmetry 
spacetime. Though the Hawking effect of Dirac particles in the 
non-stationary Kerr black hole has been studied in Ref. \cite{LZ}, the 
location and the temperature of the event horizon they obtained are only 
valid for the case of slow rotation and small evaporation\cite{ZH}, because 
their ansatz of separating variables (11) to equation (10) in Ref. \cite{LZ}, 
strictly speaking, is invalid in the most general case.  

In this letter, we try to tackle with the thermal radiation of Dirac particles 
in the non-stationary Kerr space-time. It is shown that the location and the 
temperature of event horizon is just the same as that obtained in the case of 
the thermal radiation of Klein-Gordon scalar field in the non-stationary Kerr 
space-time, but the Fermionic spectrum of Dirac particles displays another new 
effect dependent on the interaction between the spin of Dirac particles and 
the angular momentum of black holes. This has not been reported in the
previous studies yet.

The variable mass Kerr solution\cite{GHJ,CKC} can be written in the advanced 
Eddington-Finkelstein system as
\begin{eqnarray}
ds^2 &=& \frac{\Delta -a^2\sda}{\Sigma}dv^2 -2dvdr +2a\sda drd\varphi \nn\\  
&&- ~\Sigma d\theta^2 +2\frac{r^2 +a^2 -\Delta}{\Sigma}a\sda dvd\varphi \\ 
&&- ~\frac{(r^2 +a^2)^2 -\Delta a^2\sda}{\Sigma}\sda d\varphi^2 \, , \nn
\end{eqnarray}
where $\Delta = r^2 -2M(v)r +a^2$, $\Sigma = \bar{\rho}\rho$, in which $\rho 
= r -ia\cta$, $\bar{\rho}$ is the complex conjugate of $\rho$, and $v$ is the 
standard advanced time. The mass $M$ depends on the time $v$, but the specific angular 
momentum $a$ is a constant.

With the choice of a complex null-tetrad such that its corresponding 
directional derivatives are 
\begin{eqnarray}
&&\underline{D} = -\pr \, , ~~\underline{\Delta} 
= \Sigma^{-1}\left[(r^2 +a^2)\pv +2^{-1}\Delta\pr +a\pvi\right] \, , \nn\\ 
&&\delta = \frac{1}{\sqd\bar{\rho}}\left(ia\sta \pv +\pta 
+\frac{i}{\sta}\pvi\right) 
\end{eqnarray}
and $\overline{\delta}$ the complex conjugate of $\delta$, one can obtain 
the following Dirac equation\cite{Ch} 
\begin{equation}
\begin{array}{lr}
&-\left(\pr +\frac{r}{\Sigma}\right)F_1 +\frac{1}{\sqd\rho}\left({\cL} 
-\frac{ira\sta}{\Sigma}\right)F_2 = \frac{i\mu_0}{\sqd}G_1 \, , \\
&\frac{\Delta}{2\Sigma}\left({\cD} -\frac{ia\cta}{\Sigma}\right)F_2 
+\frac{1}{\sqd\bar{\rho}}\left({\bL} -\frac{a^2\sta\cta}{\Sigma}\right)F_1
= \frac{i\mu_0}{\sqd}G_2 \, , \\
&-\left(\pr +\frac{r}{\Sigma}\right)G_2 -\frac{1}{\sqd\bar{\rho}}\left({\bL} 
+\frac{ira\sta}{\Sigma}\right)G_1 = \frac{i\mu_0}{\sqd}F_2 \, , \\
&\frac{\Delta}{2\Sigma}\left({\cD} +\frac{ia\cta}{\Sigma}\right)G_1 
-\frac{1}{\sqd\rho}\left({\cL} -\frac{a^2\sta\cta}{\Sigma}\right)G_2
= \frac{i\mu_0}{\sqd}F_1 \, , \label{DCP}
\end{array}
\end{equation} 
where $\mu_0$ is the mass of Dirac particles, the operators have been 
defined as 
\begin{eqnarray*}
&{\cD}& = \pr +\Delta^{-1}\left[r -M +2a\pvi +2(r^2 +a^2)\pv\right] \, , \\ 
&{\cL}& = \pta +\frac{1}{2}\coa -\frac{i}{\sta}\pvi -ia\sta\pv \, , \\
&{\bL}& = \pta +\frac{1}{2}\coa +\frac{i}{\sta}\pvi +ia\sta\pv \, .
\end{eqnarray*}

By substituting
$$F_1 = \frac{P_1}{\sqrt{2\Sigma}} \, ,
~F_2 = \frac{\rho P_2}{\sqrt{\Sigma}} \, ,
~G_1 = \frac{\bar{\rho}Q_1}{\sqrt{\Sigma}} \, ,
~G_2 = \frac{Q_2}{\sqrt{2\Sigma}} \, , $$
into Eq. (\ref{DCP}), they have the form
\begin{equation}
\begin{array}{ll}
-\pr P_1 +{\cL} P_2 = i\mu_0\bar{\rho} Q_1 \, , 
&\Delta {\cD} P_2 +{\bL} P_1 = i\mu_0\bar{\rho} Q_2 \, ,\\
-\pr Q_2 -{\bL} Q_1 = i\mu_0\rho P_2 \, , 
&\Delta {\cD} Q_1 -{\cL} Q_2 = i\mu_0\rho P_1 \, . \label{reDP}
\end{array}
\end{equation}

Eq. (\ref{reDP}) can not be decoupled except in the case of Kerr black hole 
($M=$ const)\cite{Ch}. However, to deal with the problem of Hawking radiation, 
one may concern about the behavior of Eq. (\ref{reDP}) near the horizon only. 
Now, let us make a GTCT\cite{ZD} 
\begin{equation}
\begin{array}{ll}
&r_* = r +\frac{1}{2\kH}\ln (r -r_H) \, ,\\
&v_* = v -v_0 \, ,~\theta_* = \theta -\theta_0 \, , 
\label{trans}
\end{array}
\end{equation}
where $r_H = r_H(v, \theta)$ is the location of the event horizon, and $\kH$ 
is an adjustable parameter. All parameters $\kH$, $v_0$ and $\theta_0$ are 
constant under the tortoise transformation. 

Under the transformations (\ref{trans}), Eq. (\ref{reDP}) can be reduced to 
\begin{equation}
\begin{array}{ll}
&\spr P_1 +\left(\prH -ia \sta_0 \drH \right) \spr P_2 = 0 \, , \\ 
&\left[\Delta_H -2(r_H^2+a^2) \drH\right] \spr P_2 \\
&~~~~~~~~-\left(\prH +ia \sta_0 \drH \right) \spr P_1 = 0 \, , \label{trDPP} 
\end{array}
\end{equation}
after being taken limits $r \rightarrow r_H(v_0, \theta_0)$, $v \rightarrow 
v_0$ and $\theta \rightarrow \theta_0$. In the above, we have denoted $\drH
= \p_v r_H$, $\prH = \p_{\theta} r_H$, and $\Delta_H = r_H^2 -2M(v_0)r_H 
+a^2 $. 

If the derivatives $\spr P_1$ and $\spr P_2$ in Eq. (\ref{trDPP}) are not 
equal to zero, the existence condition of non-trial solutions for $P_1$ and 
$P_2$ is that the determinant of Eq. (\ref{trDPP}) vanishes, which gives the 
following equation to determine the location of horizon  
\begin{equation}
\Delta_H -2(r_H^2 +a^2)\drH +a^2\sda_0 {\drH}^2 +{\prH}^2 = 0 \, .
\label{loca}
\end{equation}

An apparent fact is that the Chandrasekhar-Dirac equations (\ref{reDP}) could 
be satisfied by identifying $Q_1$, $Q_2$ with $\bar{P}_2$, $-\bar{P}_1$, 
respectively. So one may deal with a pair of components $P_1$, $P_2$ only. 
Now let us consider the asymptotic behaviors of the second-order equations 
for the two-component spinor ($P_1, P_2$) near the event horizon. Given the 
GTCT in Eq. (\ref{trans}), the limiting form of the second order equations 
for $P_1, P_2$, when $r$ approaches $r_H(v_0, \theta_0)$, $v$ goes to $v_0$ 
and $\theta$ goes to $\theta_0$, reads
\begin{eqnarray}
&&{\cK}P_1 -\{r_H(1 -3\drH) -M +\pprH +\ddrH a^2\sda_0 \nn\\
&&~+~\prH \coa_0 +\frac{\Delta_H -\drH (r_H^2 +a^2) 
+ia\sta_0 \prH}{\bar{\rho}_H} \}\spr P_1 \nn\\
&&~+~\{2i\dot{M}r_H a\sta_0 -\frac{2ia\sta_0 \drH 
(r_H^2 +a^2)}{\bar{\rho}_H} \nn\\
&&~-~\frac{\Delta_H [\prH -ia\sta_0 
(\drH +1)]}{\bar{\rho}_H}\}\spr P_2 = 0 \, , \label{wone}
\end{eqnarray}
and
\begin{eqnarray}
&&{\cK}P_2 -\{ M -r_H(1 +\drH) +\pprH +\ddrH a^2\sda_0  \nn\\
&&~+~\prH \coa_0 +\frac{\Delta_H -\drH (r_H^2 +a^2) 
+ia\sta_0 \prH}{\bar{\rho}_H} \}\spr P_2 \nn\\
&&~+~\frac{\prH +ia\sta_0 (\drH -1)}{\bar{\rho}_H} \spr P_1 = 0 \, . 
\label{wtwo} 
\end{eqnarray}
where $\bar{\rho}_H = r_H +ia\cta_0$. We have used the L' H\^osptial rule
to get a term involving the second-order derivatives denoted by the operator
\begin{eqnarray*} 
{\cK} &=& \left\{\kH^{-1}[r_H(1 -2\drH) -M] -2\drH (r_H^2 +a^2) \right.\\
&&\left.~+~2\Delta_H \right\}\spdr +2(r_H^2 +a^2 -\drH a^2\sda_0) \spdvr \\
&&~+~2a(1 -\drH) \spdri P_2 -2\prH \spdra 
\end{eqnarray*}

One can adjust the parameter $\kH$ such that it satisfies
\begin{equation}
\begin{array}{ll}
&\kH^{-1}[r_H(1 -2\drH) -M] +2\Delta_H -2\drH (r_H^2 +a^2) \\
&~\equiv~ r_H^2 +a^2 -\drH a^2\sda_0 \, ,
\end{array}
\end{equation}
which means the temperature of the horizon is
\begin{equation}
\kH = \frac{r_H(1 -2\drH) -M}{r_H^2 +a^2 -\drH a^2\sda_0 
-2\Delta_H +2\drH (r_H^2 +a^2)} \, . \label{temp}
\end{equation}

Using Eq. (\ref{trDPP}), Eq. (\ref{wone}) and Eq. (\ref{wtwo}) can be 
recast into the standard wave equation near the horizon in an united form
\begin{eqnarray}
&&\spdr \Psi +2\spdvr \Psi +2\Omega \spdri \Psi +2C_3 \spdra \Psi  \nn\\
&&~~~~~~~+~2(C_2 +iC_1) \spr \Psi = 0 \, ,  \label{wave}
\end{eqnarray}
where $\Omega = a(1 -\drH)/(r_H^2 +a^2 -\drH a^2\sda_0)$ is the angular 
velocity of the event horizon of the evaporating Kerr black hole. $C_3 = 
-\prH/(r_H^2 +a^2 -\drH a^2\sda_0)$, $C_2$ (omitted here) and $C_1$ are 
all real constants, 
\begin{eqnarray}
C_1 &=& \frac{1}{2(r_H^2 +a^2 -\drH a^2\sda_0)}[-\drH a\cta_0 \nn\\
&&~+~\frac{2\dot{M}r_H\prH a\sta_0}{\Delta_H -2\drH (r_H^2+a^2)}] \, , 
~~~~{\rm for}~~\Psi = P_1, \\
C_1 &=& \frac{\drH a\cta_0}{2(r_H^2 +a^2 -\drH a^2\sda_0)} \, ,
~~~~{\rm for}~~\Psi = P_2.
\end{eqnarray}
 
Now separating as $\Psi = R(r_*)\exp[\lambda \theta_* 
+i(m\varphi -\omega v_*)]$, one has
\begin{equation}
\begin{array}{ll}
& R^{\prime\prime} +2(C_0 +iC_1 +im\Omega -i\omega)R^{\prime} = 0 \, , \\
& R \sim e^{2i(\omega -m\Omega - C_1)r_* -2C_0r_*} \, ;~~ R_0 \, . 
\end{array}
\end{equation}
where $C_0 = \lambda C_3 +C_2$, $\lambda$ is a real constant introduced
in the separation of variables. 

The ingoing wave solution and the outgoing wave solution to Eq. (\ref{wave}),
are respectively,
\begin{equation}
\begin{array}{ll}
&\Psi_{\rm in} = \exp[-i\omega v_* +im\varphi +\lambda \theta_*] \, , \\
&\Psi_{\rm out}= \Psi_{\rm in} 
e^{2i(\omega -m\Omega - C_1)r_* -2C_0r_*} \, , ~~(r > r_H) \, . 
\end{array}
\end{equation}

The outgoing wave $\Psi_{\rm out}$ is not analytic at the event horizon
$r = r_H$, but can be analytically continued from the outside of the hole 
into the inside of the hole by the lower complex $r$-plane to 
\begin{equation}
\tilde{\Psi}_{\rm out} = \Psi_{\rm out} 
e^{i\pi C_0/\kH}e^{\pi(\omega -m\Omega - C_1)/\kH} ~~(r < r_H) \, . 
\end{equation}

The relative scattering probability at the event horizon is
\begin{equation}
\left|\frac{{\Psi}_{\rm out}}{\tilde{\Psi}_{\rm out}}\right|^2
= e^{-2\pi(\omega -m\Omega - C_1)/\kH} \, . 
\end{equation}
Following the method of Damour-Ruffini-Sannan's\cite{DRS}, the Fermionic 
spectrum of Hawking radiation of Dirac particles from the black hole 
is easily obtained
\begin{equation} 
\langle {\cal N}(\omega) \rangle = 
\frac{1}{e^{(\omega -m\Omega - C_1)/T_H } +1} \, , ~~
T_H = \frac{\kH}{2\pi} \, . \label{sptr}
\end{equation} 

Equations (\ref{loca}) and (\ref{temp}) give the location and the temperature 
of event horizon of the variable-mass Kerr black hole, which depend not only 
on the advanced time $v$ but also on the polar angle $\theta$. They are in 
accord with that obtained from Klein-Gordon field equation\cite{ZDH,LJW}. 
Eq. (\ref{sptr}) shows the thermal radiant spectrum of Dirac particles in the 
non-stationary Kerr space-time, in which a residual term $C_1$ appears. The 
difference between Bosonic spectrum and Fermionic spectrum is that $C_1$ is 
absent in the spectrum of Klein-Gordon scalar particles. Also $C_1$ vanishes 
when a black hole is stationary or has a vanishing angular momentum. 

In conclusion, the $C_1$ term represents a new effect in the spectrum of Dirac 
particles, which is absent from the spectrum of Klein-Gordon particles. This 
new effect maybe arise from the interaction between the spin of Dirac particle 
and the evaporating black hole. 

S.Q. Wu is very grateful to Dr. Jeff Zhao at Motomola Company for his longterm 
helps. This work is supported in part by the NSFC under Grant No.19875019.



\begin{references}

\bibitem{Hawk}
Hawking S W 1974 {\it  Nature} {\bf 248} 30;
1975 {\it Commun. Math. Phys.} {\bf 43} 199

\bibitem{ZD}
Zhao Z and Dai X X 1991 {\it Chin. Phys. Lett.} {\bf 8} 548

\bibitem{ZDH}
Zhao Z, Dai X X and Huang W H 1993 {\it Acta Astrophysica Sinica}
{\bf 13} 299

\bibitem{LJW}
Luo M W 2000 {\it Acta Physica Sinica} {\bf 49} 1035;
Jing J L and Wang Y J 1997 {\it Int. J. Theor. Phys.} {\bf 36} 1745

\bibitem{Ch}
Chandrasekhar S 1983 {\sl The Mathematical Theory of Black Holes} 
(New York: Oxford University Press)

\bibitem{LZ}
Li Z H and Zhao Z 1995 {\it Science in China} A {\bf 38} 74

\bibitem{ZH}
Zhao Z and Huang W H 1992 {\it Chin. Phys. Lett.} {\bf 9} 333

\bibitem{GHJ}
Gonzalez C, Herrera L and Jimenez J 1979 {\it J. Math. Phys}. {\bf 20} 837; 
Jing J L and Wang Y J 1996 {\it Int. J. Theor. Phys.} {\bf 35} 1841

\bibitem{CKC}
Carmeli M and Kaye M 1977 {\it Ann. Phys.} {\bf 103} 197;
Carmeli M 1982 {\sl Classical Fields: General Relativity and Gauge Theory}
(New York: John Wiley Sons)  
 
\bibitem{DRS} 
Damour T and Ruffini R 1976 {\it Phys. Rev}. D {\bf 14} 332;
Sannan S 1988 {\it Gen. Rel. Grav.} {\bf 20} 239

\end{references}
\end{document}